\documentclass[]{article}
\usepackage[T1]{fontenc}
\usepackage[latin9]{inputenc}
\usepackage[letterpaper]{geometry}
\usepackage{amsmath}
\usepackage{amstext}
\usepackage{amssymb}
\usepackage{graphicx}
\usepackage{setspace}
\usepackage{esint}
\usepackage[authoryear]{natbib}

\makeatletter

\providecommand{\tabularnewline}{\\}

\begin{document}

\title{Bandwidth Selection In Pre-Smoothed Particle Filters}

\author{Tore Selland Kleppe\footnote{University of Stavanger, Department of Mathematics and Natural Sciences, 4036 Stavanger, Norway.
Corresponding author: e-mail: tore.kleppe@uis.no, telephone: +4751831717, fax: +4751831750.
} \and Hans Julius Skaug\footnote{University of Bergen, Department of Mathematics, Postboks 7800, 5020 Bergen, Norway.}}


\maketitle

\begin{abstract}
For the purpose of maximum likelihood estimation of static parameters, 
we apply a kernel smoother to the particles in the standard SIR filter
for non-linear state space models with additive Gaussian observation
noise. This reduces the Monte Carlo error in the estimates of both
the posterior density of the states and the marginal density of the
observation at each time point. We correct for variance inflation
in the smoother, which together with the use of Gaussian kernels,
results in a Gaussian (Kalman) update when the amount of smoothing
turns to infinity. We propose and study of a criterion
for choosing the optimal bandwidth $h$ in the kernel smoother. 
Finally, we illustrate our approach using
examples from econometrics. Our filter is shown to be highly suited
for dynamic models with high signal-to-noise ratio, for which the
SIR filter has problems.
\end{abstract}


\section{Introduction}

State space models are commonly used to represent dynamical systems
in a wide range of scientific fields. For linear and Gaussian state
space models, the Kalman Filter can be used to sequentially obtain
the posterior mean and covariance of the current state vector, as
well as the likelihood function required for estimation of model parameters.
Gaussian mixture filters \citep{alspach_sorenson_1972} were among
the first attempts to account for non-normality in the posterior,
resulting from non-linearity, either in the state equation or in the
observation equation. Later, sequential Monte Carlo (MC) based filtering
methods, collectively known as particle filters, were introduced \citep{gordon_salmond_smith_93,kitagawa_1996,liu:chen:1998}.
The particle filter has the prospect of providing a sampling-based
consistent estimate of the posterior distribution, but in many cases
the sample size (number of particles) required to bring
the MC error within tolerable bounds is prohibitively large. Consequently,
there is now a large literature on improving the baseline particle
filtering algorithms to work for a moderate numbers of particles. These
include \citet{pitt_shepard_1999}, various methods proposed in the
chapters of \citet{doucet_etal_2001} and more recently \citet{RSSB:RSSB642},
\citet{Chorin13102009} and \citet{chopin_smc2}.


Recently, a renewed interest in the use of particle filters for computing
marginal likelihood (integrating over state variables) for the purpose
of parameter estimation has emerged \citep{fv_rr_2007,RSSB:RSSB736,kantasetal09,Malik2011,dejong_unpub}.
This is also the context of the present paper. Similar to \citet{Malik2011}
and \citet{dejong_unpub} we obtain a likelihood approximation which
is continuous in the parameters, hence facilitating numerical optimization.
We target situations with highly non-linear state evolution, high
signal-to-noise ratios, and with low-to-moderate dimensional state vector, for which
adaptation is difficult. 

Throughout, we assume that the measurement model
is linear and Gaussian, which at first glance may appear restrictive. 
However, non-linear measurement equations
with additive Gaussian noise can also be handled by a simple augmentation
of the state vector, as shown in Section~\ref{sub:The-updating-problem} below.

Let $x_{t}$ and $y_{t}$ denote the state vector and observation
vector, respectively, at time $t$, and define $Y_{t}=[y_{1},\ldots,y_{t}]$.
In particle or ensemble methods the predictive density $p\left(x_{t+1}|Y_{t}\right)$
is represented by a random sample. We use a kernel smoother
$\hat{p}\left(x_{t+1}|Y_{t}\right)$ which can be updated analytically
against a linear Gaussian measurement model $p(y_{t+1}|x_{t+1})$.
From the resulting mixture approximation of the posterior $p\left(x_{t+1}|Y_{t+1}\right)$
we draw a uniformly weighted sample of particles, which after a parallel
run through the state equations, constitutes the approximation of
the next predictive distribution $p\left(x_{t+2}|Y_{t+1}\right)$.
The resulting filter, which we call the Pre-Smoothed Particle Filter
(PSPF)  is a special case of the preregularized
particle filter \citep{leglandetal98,hurzeler1998}. 

The main contribution of the present paper is to determine the optimal
amount of smoothing in each updating step of the PSPF.  This
is done adaptively, i.e. for each time point $t$ an optimal bandwidth paramter $h$
is sought. For small $h$ the PSPF approaches the SIR filter, i.e.~has
low bias but high variance. Further, we correct for variance
inflation \citep{MC19913}, and hence when $h\rightarrow\infty$ the kernel estimate $\hat{p}\left(x_{t+1}|Y_{t}\right)$
reduces to a Gaussian density with mean and covariance calculated
from the ensemble representation of $p\left(x_{t+1}|Y_{t}\right)$. At this end of the $h$ spectrum the PSPF
is strongly related to the Ensemble Kalman Filter \citep{stordal_et_al2011},
which has low MC variance but high bias.

The rest of this paper is laid out as follows. Section \ref{sec:Notation}
introduces notation and explains challenges related to particle filtering.
Section \ref{sec:PSPF} explains the pre-smoothed update and
provides a method for automatic bandwidth selection.
Section 4 introduces the PSPF, and also compares the PSPF to other particle filters using 
simulation experiments. Finally, Section \ref{sec:Illustrations} outlines two
realistic applications, and Section \ref{sec:Discussion} provides a discussion.

\section{\label{sec:Notation}Model, Notation and Background}

\subsection{Model and notation}

We consider a generic state space model consisting of a state transition
equation and an observation equation, with the former given by 
\begin{equation}
x_{t}=g(x_{t-1},v_{t}),t=1,\ldots,T,\label{eq:gen_SS_model_trans}
\end{equation}
where $g(\cdot,v_{t})$ is the state transition function \\($\mathbb{R}^{d_{x}}\rightarrow\mathbb{R}^{d_{x}}$).
The random disturbance term $v_{t}$, which can account for incomplete
model specification, may be either absent, of fixed dimension, or
of infinite dimension in the case that the state dynamics are governed
by a stochastic differential equation. Under the assumption that the
$v_{t}$s are independent (\ref{eq:gen_SS_model_trans}) describes
a Markov process, with transition probability density denoted
by $p(x_{t}|x_{t-1})$. Given the realization of $v_{t}$, evaluation
of $g(\cdot,v_{t})$ typically amounts to solving a differential equation.
The system (\ref{eq:gen_SS_model_trans}) is initialized by drawing
$x_{0}$ from a distribution with density $p(x_{0})$. It is assumed
that $g(\cdot,\cdot)$ and $v_{t}$ are sufficiently regular to ensure that $p(x_t|x_{t-1})$ is continuous, and thereby that all involved conditional distributions can be estimated consistently using kernel density estimators.

The observation equation is 
\begin{equation}
y_{t}=\mathcal{M}x_{t}+\varepsilon_{t},\;\varepsilon_{t}\sim N(0,\Sigma_{\varepsilon}),\; t=1,\dots,T,\label{eq:gen_SS_model_obs}
\end{equation}
where $y_{t}\in\mathbb{R}^{d_{y}}$, $\Sigma_{\varepsilon}\in\mathbb{R}^{d_{y}\times d_{y}}$
is non-degenerate and the matrix $\mathcal{M}\in\mathbb{R}^{d_{y}\times d_{x}}$
is independent of the state, but may vary non-stochastically with
time $t$. Moreover, we use the notation $Y_{t}\equiv[y_{1},\dots,y_{t}]$,
$Y_{0}=\emptyset$.
$\mathcal{N}(x|\mu,\Sigma)$ denotes the multivariate Gaussian probability
density function evaluated at $x$, $I_{q}$ the $q\times q$ identity
matrix. Finally, we indicate which
stochastic variable an expectation or variance is taken over using subscripts (e.g.
$E_x$ when expectation is taken over variable $x$).

\subsection{\label{sub:The-SIR-filter}The SIR filter and sample impoverishment}

This section introduces particle filters and the Sampling Importance Resampling (SIR) filter of \citet{gordon_salmond_smith_93},
which is the limit of PSPF as $h\rightarrow 0$.
Any particle filtering approach relies on alternating between two
steps: prediction (p) in which $p(x_{t+1}|Y_{t})$ is represented
by a random sample $\{x_{t+1}^{(i),p}\}_{i=1}^{n}$, and filtering
(f) in which $p(x_{t+1}|Y_{t+1})$ similarly is approximated by $\{x_{t+1}^{(i),f}\}_{i=1}^{n}$.
These random samples of size $n$ are referred to as filter- and predictive
swarms, respectively, and are updated iteratively from each other.
The prediction step, used in both SIR and PSPF, consists of $x_{t+1}^{(i),p}=g\left(x_{t}^{(i),f},v_{t+1}^{(i)}\right),\; i=1,\dots,n$,
where the $v_{t+1}^{(i)}$ are independent random draws from the distribution
of $v_{t+1}$. In the filtering step Bayes formula is invoked: 
\begin{multline}
p(x_{t+1}|Y_{t+1})=\frac{p(y_{t+1}|x_{t+1})p(x_{t+1}|Y_{t})}{\int p(y_{t+1}|x_{t+1})p(x_{t+1}|Y_{t})dx_{t+1}}\\=\frac{p(y_{t+1}|x_{t+1})p(x_{t+1}|Y_{t})}{p(y_{t+1}|Y_{t})}.\label{eq:filter_eq}
\end{multline}
The SIR filter approximates (\ref{eq:filter_eq}) by performing a
SIR update \citep{rubin_1987}, representing $p(x_{t+1}|Y_{t+1})$
as a weighted sample with locations $\{x_{t+1}^{(i),p}\}_{i=1}^{n}$
and corresponding weights 
\begin{equation}
\frac{p(y_{t+1}|x_{t+1}^{(i),p})}{n\check{p}(y_{t+1}|Y_{t})},\; i=1,\dots,n,\label{eq:SIR_posterior}
\end{equation}
where $\check{p}(y_{t+1}|Y_{t})\equiv n^{-1}\sum_{i=1}^{n}p(y_{t+1}|x_{t+1}^{(i),p})$
is a normalizing constant. Obtaining a uniformly weighted sample $\{x_{t+1}^{(i),f}\}_{i=1}^{n}$
to complete next time-step's prediction is simply a matter of drawing
multinomially from $\{x_{t+1}^{(i),p}\}_{i=1}^{n}$ with weights (\ref{eq:SIR_posterior}).
A byproduct of the SIR filter is that the marginal likelihood of $Y_{t}$
needed for parameter estimation can be approximated as
\begin{equation}
p(Y_{t})=\prod_{t=1}^{T}p(y_{t}|Y_{t-1})\approx\prod_{t=1}^{T}\check{p}(y_{t}|Y_{t-1}),\label{eq:SIR_likelihood}
\end{equation}
for large $n$ (see e.g. \citet[Proposition 7.4.1.]{delmoral}). 

Sample impoverishment in the SIR filter occurs when, at time step
$t$, the predictive particle swarm $\{x_{t}^{(i),p}\}_{i=1}^{n}$
and the data likelihood $p(x_{t}|y_{t})\propto p(y_{t}|x_{t})$ are
poorly aligned (see e.g. \citet{pitt_shepard_1999}).
The multinomial probabilities (\ref{eq:SIR_posterior}) then become
very unevenly distributed, and the multinomial sampling will yield
many repeated particles. Over time the swarm will degenerate in the
sense that all particles can be traced back to a single particle in
the initial swarm ($t=0).$ Sample impoverishment also increases the
MC error of the likelihood estimator (\ref{eq:SIR_likelihood}). This
is likely to occur during numerical optimization of the likelihood,
when the optimization algorithm tries an infeasible parameter value
rendering the particle swarm and the data likelihood $p(y_{t}|x_{t})$
incompatible. The effect is amplified by a high signal-to-noise ratio
in the system. Numerous strategies have been proposed for aligning
(adapting) the predictive swarm to the coming observation (see e.g.
\citet{cappe_godsill_moulines_2007} for an overview), but these typically
rely on evaluation of $p(x_{t+1}|x_{t})$ (or some of the characteristics
of $p(x_{t+1}|x_{t})$) which may be costly. The PSPF, on the other hand,
avoids evaluation of $p(x_{t+1}|x_{t})$, and relies only on the ability
to simulate (\ref{eq:gen_SS_model_trans}) efficiently.

\section{\label{sec:PSPF}The Pre-Smoothing Update}
In this section we consider the pre-smoothing (PS) update, as an alternative
to \citet{rubin_1987}'s SIR update when the observation equation
is linear in the state and additively Gaussian. 
Focusing on a single updating step we can drop the index $t$ in our noation.
In a general perspective, the problem we address is the Bayesian updating problem of evaluating the posterior density $p(x|y)$ and
the marginal density $p(y)$ when the prior $\pi(x)$ is represented
by a random sample. In particular, we focus on optimal selection of the smoothing parameter in the PS update, 
with the overarching objective of producing accurate estimates of~$p(y)$. 
In Section \ref{sub:Pre-Smoothed-Particle-Filters} we again return to the filter setting.

\subsection{\label{sub:The-updating-problem}The updating problem}
Consider the evaluation of the posterior $p(x|y)$ and marginal $p(y)$,
for the model 
\begin{eqnarray}
y|x & \sim & p(y|x)=\mathcal{N}(y|\mathcal{M}x,\Sigma_{\varepsilon}),\label{eq:onestep_pycy}\\
x & \sim & \pi(x),\label{eq:onestep_px}
\end{eqnarray}
in a setting where $\pi$ is an unknown prior density, while $\mathcal{M}$
and $\Sigma_{\varepsilon}$ are given matrices. The available information
about $\pi$ is a random sample $\mathbf{x}=\left\{ x^{(i)}\right\} _{i=1}^{n}$
drawn from $\pi$. Our aim is to estimate both $p(x|y)$ and $p(y)$
for a given $y$. We denote by $\hat{\mu}_{x}$ and $\hat{\Sigma}_{x}$
the empirical mean and covariance matrix of the sample $\mathbf{x}$, respectively. 

Consider the shrunk kernel estimate \citep{MC19913,west_93} 
\begin{eqnarray}
\hat{\pi}(x)&=&\frac{1}{n}\sum_{i=1}^{n}\mathcal{N}(x|m_{i},G),\label{eq:pi_hatt}\\
 m_{i}&=&(1-b)\hat{\mu}_x+bx^{(i)}\label{def:m_i},
 \\ G&=&(1-b^{2})\hat{\Sigma}_{x}\label{def:G},
\end{eqnarray}
where the smoothing parameter $b\in[0,1]$ governs the proportions
of the total variance under $\hat{\pi}(\cdot)$ stemming from inter-kernel
variance ($b^{2}\hat{\Sigma}_{x}$) and intra-kernel variance ($G$)
in the Gaussian mixture (\ref{eq:pi_hatt}). The replacement of the
more conventional bandwidth parameter $h=\sqrt{b^{-2}-1}$ by $b$
simplifies certain expressions in what follows. The estimator (\ref{eq:pi_hatt})
avoids the ``variance inflation'' to which the standard kernel estimator
\citep{Si86} is prone, as it is easily verified that the mean and
variance under $\hat{\pi}(\cdot)$ is $\hat{\mu}_{x}$ and $\hat{\Sigma}_{x}$,
respectively. For $b$ close to 1 ($h$ close 0) $\hat{\pi}(\cdot)$
behaves as a standard kernel estimator with equally weighted point masses located at $x^{(i)},\;i=1,\dots,n$ as 
the limit. For $b\rightarrow0$ ($h\rightarrow\infty$) a Gaussian
limit is obtained, i.e. $\hat{\pi}(x)\rightarrow\mathcal{N}(x|\hat{\mu}_{x},\hat{\Sigma}_{x})$. 

By substituting $\hat{\pi}$ for $\pi$ in the Bayes rule~(\ref{eq:filter_eq}), we obtain the PS estimators 
\begin{equation}
\hat{p}(y)  =  \int\mathcal{N}(y|\mathcal{M}x,\Sigma_{y})\hat{\pi}(x)dx=\frac{1}{n}\sum_{i=1}^{n}W_{i},\label{eq:py_estimate}
\end{equation}
\begin{equation}
\hat{p}(x|y)  =  \frac{\mathcal{N}(y|\mathcal{M}x,\Sigma_{y})\hat{\pi}(x)}{\int\mathcal{N}(y|\mathcal{M}x,\Sigma_{y})\hat{\pi}(x)dx}\\
=\frac{\sum_{i=1}^{n}W_{i}\varphi_{i}(x)}{\sum_{i=1}^{n}W_{i}}=\sum_{i=1}^{n}w_{i}\varphi_{i}(x),\label{eq:pxcy_estimate}
\end{equation}
where
\begin{eqnarray*}
W_{i} & = & \mathcal{N}(y|\mathcal{M}m_{i},\Sigma_{\varepsilon}+\mathcal{M}G\mathcal{M}^{T}),\\
w_{i} & = & \frac{W_{i}}{n\hat{p}(y)},\\
\varphi_{i}(x) & = & \mathcal{N}(x|m_{i}+Q(y-\mathcal{M}m_{i}),G-Q\mathcal{M}G),\\
Q & = & G\mathcal{M}^{T}(\Sigma_{\varepsilon}+\mathcal{M}G\mathcal{M}^{T})^{-1}.
\end{eqnarray*}
In our notation we have omitted the dependence on $b$. 

\begin{table*}
\begin{tabular}{cll}
\hline 
 & Gaussian ($b\rightarrow0$) & SIR ($b\rightarrow1$)\tabularnewline
\hline 
$\hat{p}(y)$ & $\mathcal{N}(y,\mathcal{M}\hat{\mu}_{x},\Sigma_{\varepsilon}+\mathcal{M}\hat{\Sigma}_{x}\mathcal{M}^{T})$ & $n^{-1}\sum\mathcal{N}(y|\mathcal{M}x^{(i)},\Sigma_{\varepsilon})$\tabularnewline
$\hat{p}(x|y)$ & $\mathcal{N}(x,\hat{\mu}_{x}+\hat{K}(y-\mathcal{M}\hat{\mu}_{x}),\hat{\Sigma}_{x}-\hat{K}\mathcal{M}\hat{\Sigma}_{x})$ & $c^{-1}\sum_{i=1}^{n}\mathcal{N}(y|\mathcal{M}x^{(i)},\Sigma_{\varepsilon})\delta(x-x^{(i)})$\tabularnewline
$E(x|y)$ & $\hat{\mu}_{x}+\hat{K}(y-\mathcal{M}\hat{\mu}_{x})$ & $c^{-1}\sum_{i=1}^{n}x^{(i)}\mathcal{N}(y|\mathcal{M}x^{(i)},\Sigma_{\varepsilon})$\tabularnewline
Property & High bias, low variance & Low bias, high variance\tabularnewline
\hline 
\end{tabular}\caption{\label{tab:Properties-of-the}Limit cases ($b\rightarrow 0,1$) for the PS updating step,
where $\hat{K}=Q|_{b=0}=\hat{\Sigma}_{x}\mathcal{M}^{T}(\Sigma_{\varepsilon}+\mathcal{M}\hat{\Sigma}_{x}\mathcal{M}^{T})^{-1}$
is the Kalman gain matrix, $\delta(x)$ denotes a unit point mass located at the origin and $c=\sum_{i=1}^{n}\mathcal{N}(y|\mathcal{M}x^{(i)},\Sigma_{\varepsilon})$
is a normalizing constant. }
\end{table*}

As $b$ varies from 1 to 0 the PS updates moves from a SIR update
to the Gaussian update, both of which are summarized in Table \ref{tab:Properties-of-the}.
The mean $m_{i}+Q(y-\mathcal{M}m_{i})$ of each posterior mixture
component concurrently moves smoothly from $x^{(i)}$ ($=x^{(i),p}$) toward what
is dictated by the likelihood, reducing the potential for sample impoverishment.
In the same vein, we have $w_{i}\rightarrow n^{-1}$ as $b\rightarrow0$,
i.e.~uniform weighting. These properties of the PS update (and the updates employed
in other pre-regularized filters) differ from those
of the update mechanisms employed in post-smoothed particle filters
advocated by \citet{musso01improving} and \citet{Flury09learningand},
where the (one step) posterior locations and weights are unchanged
relative to the SIR. However, these latter approaches do not require the
Gaussian data-likelihood which is underlying the PS update. 

The fact that $\hat{p}(x|y)$ is a finite Gaussian mixture (for $b<1$)
has a number practical advantages. Firstly, moments, marginal- and
conditional distributions of the approximate posterior are easily
derived from the representation (\ref{eq:pxcy_estimate}). Further,
$\hat{p}(\cdot|y)$ has continuous support, and therefore direct copying
of particles, which is applied in the SIR filter, is avoided in the
resampling step. Sampling from $\hat{p}(\cdot|y)$ is trivial. Moreover
continuous (with respect to the parameters) sampling, resulting
in a continuous simulated likelihood function, can be implemented.

The apparently restrictive linearity assumption~(\ref{eq:onestep_pycy}) can be relaxed 
by augmenting the state variable $x$. The case of a non-linear measurement function 
$M(x)$ with additive Gaussian noise can be accommodated
without any conceptual change to the framework. The measurement variance $\Sigma_{\varepsilon}$ 
is then split in two parts $r^2\Sigma_{\varepsilon}$ and $(1-r^2)\Sigma_{\varepsilon}$, $0<r<1$,
and the augmented state vector is $x^{\prime} = [x,M(x)+\eta]^T$
where $\eta \sim N(0,r^2\Sigma_{\varepsilon})$ is an auxiliary variable introduced for convenience.
For the augmented system, equations (\ref{eq:onestep_pycy})-(\ref{eq:onestep_px}) take the form
\begin{eqnarray*}
y&\sim& N(\mathcal M^{\prime} x^{\prime},(1-r^2)\Sigma_{\varepsilon}),\\
x^{\prime}&\sim&\pi^{\prime}(x^{\prime}),
\end{eqnarray*}
where $\pi^{\prime}$ is the induced prior and $\mathcal M$ is the matrix 
that selects $M(x)+\eta$ from $x^{\prime}$. Now, $r$ is a tuning parameter
that must be chosen jointly with $b$. Estimates of $p(x|y)$ are easily
obtained as a marginal in the finite mixture representation of $\hat p(x^{\prime \prime}|y)$.
An application of this approach is given in section~\ref{sub:Example:-DSGE-model} below.

\subsection{Criterion for smoothing parameter selection\label{sec:pycriterion}}

A critical part of the PS update is the selection of the smoothing parameter,
with the aim of obtaining both a representative posterior particle swarm and
and accurate estimate $\hat p (y)$ of the marginal likelihood. For this
purpose \citet{Flury09learningand} argue that the integrated mean
squared error (MISE) of $\hat{p}(x|y)$, which is commonly used in the kernel
smoothing literature \citep{Si86} is not a suitable criterion in a particle 
filter setting. Nevertheless, is has been used for pre- and post-smoothed 
particle filters by e.g. \cite{leglandetal98,hurzeler1998,musso01improving}.
Instead, \citet{Flury09learningand}
propose to minimize the MISE of the posterior cumulative distribution
function. We propose a third criterion, namely to minimize the
mean squared error (MSE) of $\hat{p}(y)$, which is given as
\begin{eqnarray}
MSE(\hat{p}(y)) & = & (E_{\mathbf x}(\hat{p}(y))-p(y))^{2}+Var_{\mathbf x}(\hat{p}(y)) \nonumber \\
 & \equiv & C(b).\label{eq:MSE(py)}  
\end{eqnarray}
This criterion has the advantage of being analytically simple, in addition
to targetting minimal Monte Carlo error in the likelihood function as explained below.
Minimization of $C(b)$ gives an optimal bias-variance balance
that depends on the observation $y$. 

Switching momentarily to a dynamical system setting (with $\mathbf{x}=\{x_{t}^{(i),p}\}_{i=1}^{n}$)
for the reminder of this paragraph,
$\hat{p}(y)$ estimates $p(y_{t}|Y_{t-1})$, 
 and thus choosing (\ref{eq:MSE(py)})
as the criterion targets directly the factors involved in 
the likelihood function (\ref{eq:SIR_likelihood}).
However, it should be noted in the dynamic setting that
current period's filter distribution must be represented accurately as it
serves as an important input to next period's likelihood evaluation. 
We show in section~\ref{sub:Comparison-with-other} that using an approximation to
$C$ (which targets $p(y)$) 
also leads to competitive estimates of the filtering distribution 
(i.e. $p(x|y)$). This will in particular be true whenever most of the
information carried in $p(x|y)$ comes from the likelihood (which is typical
for the class of models we consider) as $\hat p(x|y)$ is almost proportional 
to $x\mapsto p(y|x)$, and therefore the posterior estimator is relatively 
insensitive to the choice of smoothing parameter. On the other hand, assuming 
a concentrated observation likelihood, and in addition that $\mathcal{M}$ is 
invertible (i.e. $d_x=d_y$), $\hat p(y)$ will be highly sensitive to the choice
of smoothing parameter since a zeroth order approximation of $\hat p(y)$ is 
proportional to $\hat \pi(\mathcal{M}^{-1} y)$. Hence it appears sensible to choose $C$ even 
in a dynamic setting, in particular in high signal-to-noise situations.

\subsection{Plug-in and approximation\label{sub:Operational-smoothing-parameter}}
There are two obstacles to direct use of the criterion $C(b)$ as given 
in~(\ref{eq:MSE(py)}).  First, the expectation is taken over~$\mathbf x$ 
which has unknown distribution $\pi$ (see~(\ref{eq:onestep_px})). 
The same problem occurs in standard
kernel estimation, and is solved by the use of a plug-in estimator~\citep{Si86}. 
The second problem is caused by the use of the shrunk kernel estimator,
which involves the empirical (depending on~$\mathbf x$) quantities 
$\hat{\mu}_{x}$ and $\hat{\Sigma}_{x}$ through~(\ref{def:m_i}) and~(\ref{def:G}). 
Even if $\pi$ was known, analytical evaluation of the expectation in~(\ref{eq:MSE(py)})
would not be possible, and we have to resort to an approximation. 
\citet{MC19913} encounters the same problem, but argues that the
effect can be ignored asymptotically when $n\rightarrow\infty$ and $b\rightarrow 1$.
However, we consider the full range $b\in(0,1)$ so the
same asymptotic arguments do not apply. Instead we
attempt to approximate the expectation~(\ref{eq:MSE(py)}) for finite~$n$.  

We start by addressing the second problem. For the purpose of 
evaluating the mean and variance in~(\ref{eq:MSE(py)}) we replace 
$\hat{\mu}_{x}$ and $\hat{\Sigma}_{x}$ in expressions 
(\ref{eq:pi_hatt}-\ref{eq:py_estimate}) by
new random variables, $\tilde\mu$ and $\tilde\Sigma$, respectively,
both taken to be independent of $\mathbf x$. The simplification (approximation)
lies mostly in this independence assumption, but also in the distributional assumptions
made about $\tilde\mu$ and $\tilde\Sigma$ below.
The reason that we cannot ignore the sampling variability 
in $\hat \mu_x$ and $\hat \Sigma_x$ is that the variance term in (\ref{eq:MSE(py)}) 
would then be exactly zero for $b=0$.
Hence, for small $b$ we would underestimate the MSE of $\hat{p}(y)$.

We make the following distributional choices
\begin{equation}
\tilde\mu\sim N(\mu_x,\Sigma_x/n),   \label{def:tilde_mu}
\end{equation}
and \begin{equation}
\tilde\Sigma\sim \frac{1}{n}\text{Wishart}(\Sigma_x,n-1),  \label{def:tilde_Sigma}
\end{equation}
i.e.~$\tilde\mu$ and $\tilde\Sigma$ are distributed as if they where 
calculated from a sample of $n$ iid $N(\mu_x, \Sigma_x)$ vectors. 
Plug-in versions of (\ref{def:tilde_mu}) and (\ref{def:tilde_Sigma}),
i.e.~where $\mu_x$ has been replaced by $\hat \mu_x$ and $\Sigma_x$ by $\hat\Sigma_x$,
are used immediately below for notational convenience. 
Strictly speaking these replacements take place after all 
moment calculations have been carried out.

After these simplifications, it is necessary to restate our criterion
\begin{equation}
\tilde C(b)  =  \left[E(\tilde{p}(y))-p(y)\right]^{2}
	+Var(\tilde{p}(y))\label{eq:C_tilde} 
\end{equation}
where expectation and variance now is taken relative to $\mathbf x$,
$\tilde\mu$ and $\tilde\Sigma$,
which we emphasize are independent by assumption.
Writing out the details of~(\ref{eq:C_tilde}) we get $\tilde{p}(y)  =  n^{-1}\sum_{i=1}^{n}\tilde W_{i}$
where
\begin{equation}
\tilde W_i = \mathcal N \{y|\mathcal M (a\tilde\mu+bx^{(i)}),\\
\Sigma_\varepsilon + G^\prime \mathcal M \tilde\Sigma \mathcal M^T\},
\end{equation}
with $a\equiv 1-b$ and $G^\prime \equiv 1-b^2$.


The next sections outline pilot distributions and 
develop asymptotic approximations (in $n$) that will enable us to evaluate the mean and variance in~(\ref{eq:C_tilde}).
\begin{table*}
\begin{tabular}{ll}\hline
Expression & Interpretation\\ \hline \\
$f_0(\tilde\Sigma) 
=\sum_{l=1}^{2}\hat{q}_{l}
\mathcal{N}\Big(y|a\mathcal{M}\hat{\mu}_{x}+b\mathcal{M}\hat{\mu}_{l},
v_l \Big)$, &
$\underset{\tilde\mu}{E}\left[\underset{x\sim \hat{\pi}_{B}}{E}
\left( \tilde W |\tilde\mu,\tilde\Sigma \right)| \tilde\Sigma \right]$.\\
where $v_l = \Sigma_{\varepsilon}+b^{2}\mathcal{M}\hat{\Sigma}_{l}\mathcal{M}^T+
\frac{a^{2}}{n}\mathcal{M}\hat{\Sigma}_{x}\mathcal{M}^{T}
+G^{\prime}\mathcal{M}\tilde{\Sigma}\mathcal{M}^{T}$.\\ \\ \hline 
\\
$f_{1}(\tilde\Sigma) = \mathcal{N}(y|\mathcal{M}\hat{\mu}_{x},
 \Sigma_{\varepsilon}+(b^{2}+a^{2}/n)\mathcal{M}\hat{\Sigma}_{x}\mathcal{M}^{T}
 +G^{\prime}\mathcal{M} \tilde \Sigma \mathcal{M}^{T})$. 
 & $\underset{\tilde\mu}{E}\left[
\underset{x\sim\hat{\pi}_{V}}{E}\left(\tilde W|\tilde\mu,\tilde\Sigma \right) | \tilde\Sigma\right]$.\\
\\
$f_{2}(\tilde\Sigma) = \frac{\mathcal{N}\left(y|\mathcal{M}\hat{\mu}_{x},
\frac{1}{2}\Sigma_{\varepsilon}+\left(b^{2}+\frac{a^{2}}{n}\right)
\mathcal{M}\hat{\Sigma}_{x}\mathcal{M}^{T}
+\frac{G^{\prime}}{2}\mathcal{M}\tilde{\Sigma}\mathcal{M}^{T} \right)}
{(4\pi)^{d_{y}/2}\sqrt{|\Sigma_{\varepsilon}+G^{\prime}\mathcal{M}\tilde{\Sigma}\mathcal{M}^{T}|}}$.
 & $\underset{\tilde\mu}{E}\left[
\underset{x\sim\hat{\pi}_{V}}{E}\left(\tilde W^2|\tilde\mu,\tilde\Sigma \right) | \tilde\Sigma\right]$.\\ 
\\
$f_{3}(\tilde\Sigma) = \frac{\mathcal{N}\left(y|\mathcal{M}\hat{\mu}_{x},\frac{1}{2}\Sigma_{\varepsilon}+\left(\frac{b^2}{2}+\frac{a^{2}}{n}\right)
\mathcal{M}\hat{\Sigma}_{x}\mathcal{M}^{T}+
\frac{G^\prime}{2}\mathcal{M}\tilde{\Sigma}\mathcal{M}^{T}\right)}
{(4\pi)^{d_{y}/2}\sqrt{|\Sigma_{\varepsilon}+b^2\mathcal{M}\hat{\Sigma}_{x}\mathcal{M}^{T}+G^\prime \mathcal{M}\tilde{\Sigma}\mathcal{M}^{T}|}}.$ 
& $\underset{\tilde\mu}{E}\left[
\left[ \underset{x\sim\hat{\pi}_{V}}{E}\left(\tilde W|\tilde\mu,\tilde\Sigma \right) \right]^2 | \tilde\Sigma\right].$ \\  \\ \hline \\
$\breve{f}_{1} = \left[\left(\mathcal{M}^{T}F^{-1}\bar{y}\right)\left(\mathcal{M}^{T}F^{-1}\bar{y}\right)^{T}-\mathcal{M}^{T}F^{-1}\mathcal{M}\right]$, &
$\frac{2}{G^{\prime}}\nabla_{\tilde{\Sigma}}\log f_{1}(\tilde{\Sigma})|_{\tilde{\Sigma}=\hat \Sigma_{x}}$.\\
where $F = \Sigma_{\varepsilon}+(1+a^{2}/n)\mathcal{M}\hat{\Sigma}_{x}\mathcal{M}^{T}$
and $\bar{y}\equiv y-\mathcal{M}\hat{\mu}_{x}$. \\
\\
\hline
\end{tabular}
\caption{\label{tab:MSEexpr} Expressions used for calculating the approximate MSE $\tilde C(b)$. The calculations
leading to $f_0,f_1,f_2,f_3$ are tedious but trivial, as they only involve integration over (unnormalized) multivariate
normal distributions.}
\end{table*}

\subsubsection{Pilot distributions}
%
For the variance term in $\tilde C$, we employ for convenience a Gaussian pilot,
\begin{equation}
\hat{\pi}_{V}(x)\equiv\mathcal{N}(x|\hat{\mu}_{x},\hat{\Sigma}_{x}).
\end{equation}
For the squared bias term in~(\ref{eq:C_tilde}) a Gaussian pilot is ruled out because, as shown below,
this leads asymptotically to zero bias for all $b$. Instead a two-component Gaussian mixture
\begin{equation}
\hat{\pi}_{B}(x)\equiv\sum_{l=1}^{2}\hat{q}_{l}\mathcal{N}(x|\hat{\mu}_{l},\hat{\Sigma}_{l}),
\end{equation}
is used. The bias pilot $\hat{\pi}_{B}$ is flexible, allowing for analytical
computations of moments, and \\ $\{\hat{q}_{l},\hat{\mu}_{l},\hat{\Sigma}_{l}\}_{l=1}^{2}$
may be estimated from $\mathbf{x}$ using an EM-algorithm \citep[see e.g.][section 2.8 for details]{mclachlan2000fmm}.
To minimize the computational burden we perform only a few EM-iterations,
and further computational savings are obtained by running the EM on
a subsample of $\mathbf{x}$ when $n$ is large.

\subsubsection{Practical squared bias}
Under the above introduced simplifying 
approximations, and in particular under pilot density $\hat \pi_B$, we have that
\begin{eqnarray}
E(\tilde p(y))&= & \underset{\tilde\Sigma,\tilde\mu,x^{(i)}\sim \text{ iid }\hat{\pi}_{B}}{E}\left(\frac{1}{n}\sum_{i=1}^n \tilde W_i\right)\nonumber\\
& =&\underset{\tilde\Sigma}{E}\left[\underset{\tilde\mu}{E}\left[ \underset{x\sim \hat{\pi}_{B}}{E}
\left( \tilde W |\tilde\mu,\tilde\Sigma \right)| \tilde\Sigma \right]\right] \nonumber \\
&=&\underset{\tilde\Sigma}{E}\left[f_0(\tilde \Sigma)\right]. \label{eq:wtildemean}
\end{eqnarray}
A closed form expression for $f_0(\tilde \Sigma)$ is given in Table~\ref{tab:MSEexpr}.
The expectation over $\tilde\Sigma$ in (\ref{eq:wtildemean}) does not appear to have closed form, and we therefore 
employ the asymptotical (in $n$) mean statement of Corollary 2.2 of \citet{iwashita_siotani_94}
to obtain
\begin{equation} 
E(\tilde p(y))=\underset{\tilde\Sigma}{E}\left[f_0(\tilde \Sigma)\right] 
  \simeq f_0(\hat \Sigma_x) 
  \equiv \hat \rho_B(b;y),\label{eq:finalrhobhat} 
\end{equation}
where $\hat \rho_B$ serves as the practical approximation to $E(\tilde p (y))$. 

Note that $p(y)=E_x[p(y|x)]$ in (\ref{eq:tildeC}) depends on $\pi$
and is hence estimated using the pilot density needs to be estimated. 
Under the pilot density $\hat{\pi}_{B}$ it 
has a closed form expression
\begin{eqnarray}
E_x[p(y|x)]&\approx & \underset{x \sim \hat{\pi}_{B}}{E}\left[ \mathcal N (y|\mathcal M x, \Sigma_\varepsilon) \right],\nonumber\\
& = & \sum_{l=1}^{2}\hat{q}_{l}\mathcal{N}(y|\mathcal{M}\hat{\mu}_{l},\Sigma_{\varepsilon}+\mathcal{M}\hat{\Sigma}_{l}\mathcal{M}^{T}),\nonumber\\
& \equiv & \rho_B(b;y) \label{eq:rhoB}.
\end{eqnarray}
Finally, $\left(\hat \rho_B - \rho_B\right)^2$ is taken as the practical squared bias term. 
Note in particular that it is easily verified that also the practical squared
bias vanishes for $b=1$, as $a=G^\prime =0$ in this case.

To underpin the claim that a non-Gaussian bias pilot is needed, 
we momentarily choose the parameters of $\hat \pi_B$ so that $\hat \pi_B$ 
coincides with $\hat \pi_V$, 
e.g. via $q_1=1$, $q_2=0$, $\hat \mu_1=\hat \mu_x$, $\hat \Sigma_1=\hat \Sigma_x$. Then   
$\hat \rho_B=\mathcal N(y|\mathcal M \hat \mu_x, \Sigma_\varepsilon+(1+a^2/n)\mathcal{M}\hat{\Sigma}_{x}\mathcal{M}^T)$
whereas
$\rho_B=\mathcal N(y|\mathcal M \hat \mu_x, \Sigma_\varepsilon+\mathcal{M}\hat{\Sigma}_{x}\mathcal{M}^T),$
which shows that the practical bias would vanish as $n\rightarrow\infty$ for all $b$ if a 
Gaussian bias pilot was employed.
\subsubsection{Practical variance}

The variance of $\tilde p(y)$, taken under pilot density $\hat{\pi}_{V}$,
relies on the identity developed in Appendix \ref{sec:operational_calc}:
\begin{eqnarray}
& & Var(\tilde p(y)) \nonumber\\
&=& \underset{\tilde\Sigma,\tilde\mu,x^{(i)}\sim \text{ iid }\hat{\pi}_{V}}{Var}\left(\frac{1}{n}\sum_{i=1}^n \tilde W_i\right) \nonumber \\
&=& \underset{\tilde\Sigma}{Var}(f_1(\tilde\Sigma))
 + \underset{\tilde\Sigma}{E}(f_3(\tilde\Sigma))
 - \underset{\tilde\Sigma}{E}(f_1(\tilde\Sigma)^2) \nonumber \\
& &+\frac{1}{n}\left(\underset{\tilde\Sigma}{E}(f_2(\tilde\Sigma)) 
- \underset{\tilde\Sigma}{E}(f_3(\tilde\Sigma)) \right),\label{eq:MSE_var_eq}
\end{eqnarray}
where explicit expressions for $f_1,f_2,f_3$ are  can be found in Table \ref{tab:MSEexpr}. As in the
calculations leading to the squared bias, the expectations and variance 
 over $\tilde\Sigma$ in (\ref{eq:MSE_var_eq}) do not appear to have closed form expressions.
Consequently we employ mean statement of Corollary 2.2 of \citet{iwashita_siotani_94} to obtain  
\begin{eqnarray}
& &\underset{\tilde\Sigma}{E}(f_3(\tilde\Sigma))
 - \underset{\tilde\Sigma}{E}(f_1(\tilde\Sigma)^2) \nonumber \\
& &+\frac{1}{n}\left(\underset{\tilde\Sigma}{E}(f_2(\tilde\Sigma)) 
- \underset{\tilde\Sigma}{E}(f_3(\tilde\Sigma)) \right) \nonumber \\
&\simeq& f_3(\hat \Sigma_x) - f_1(\hat \Sigma_x)^2 + \frac{1}{n}
\left(f_2(\hat \Sigma_x)-f_3(\hat \Sigma_x)\right) \nonumber \\
&\equiv& \rho_{V,1}(y;b).\label{eq:pracV1}
\end{eqnarray}
The variance of $f_1(\tilde\Sigma)$ in (\ref{eq:MSE_var_eq}) is treated 
using the delta rule variance statement of Corollary 2.2 of \citet{iwashita_siotani_94}:
\begin{eqnarray}
& &\underset{\tilde\Sigma}{Var}(f_1(\tilde\Sigma)) \simeq
\frac{1}{2n}f_{1}(\hat \Sigma_{x})^{2}\left(G^{\prime}\right)^{2}\text{tr}\left[(\breve{f}_{1} \hat \Sigma_{x})^{2}\right]\nonumber \\ 
&\equiv& \rho_{V,2}(y;b)\label{eq:pracV2}.
 \end{eqnarray}
Here, $\breve{f}_{1}$ is the Jacobian matrix of $\log f_1(\tilde\Sigma)$ up to a factor 
$2/G^\prime$ with
explicit expression given in Table \ref{tab:MSEexpr}, and
$\text{tr}$ denotes the matrix trace. Combining (\ref{eq:pracV1}) and (\ref{eq:pracV2}) with
(\ref{eq:MSE_var_eq}) yields the practical variance approximation $\rho_V$ as
\begin{equation}
Var\left(\tilde p(y)\right) \simeq \rho_{V,1}+\rho_{V,2}\equiv \rho_V.\label{eq:finalval} \\
\end{equation}
Finally, collecting the results obtained in (\ref{eq:finalrhobhat}), (\ref{eq:rhoB}) and (\ref{eq:finalval}) 
yields the practical MSE approximation
\begin{equation}
\bar C(b) \equiv \left(\hat \rho_B(b)-\rho_B(b)\right)^2+\rho_V(b), \label{eq:tildeC}
\end{equation}
which will be used throughout the rest of the paper. 

\subsubsection{Implementation\label{sec:bwimpl}}
Given a prior sample $\mathbf x$, observation $y$ and matrices $\mathcal M$, 
$\Sigma_\varepsilon$, the location of one approximately optimal 
smoothing parameter $\bar b$ involves the following steps:
\begin{enumerate}
\item Calculate $\hat \mu_x$, $\hat \Sigma_x$ from $\mathbf x$ and estimate
$\{\hat{q}_{l},\hat{\mu}_{l},\hat{\Sigma}_{l}\}_{l=1}^{2}$
from (possibly a subset of) $\mathbf{x}$ using a few iterations of
the EM-algorithm.
\item Compute $\rho_B$ using (\ref{eq:rhoB}).
\item Numerically minimize (\ref{eq:tildeC}) with respect to $b\in[0,1]$, 
and return the minimizer as $\bar b$.
\end{enumerate}
Our MATLAB implementation of the smoothing parameter selection procedure uses
the built in function gmdistribution.fit for estimating $\hat \pi_B$
and the minimization of $\bar C(b)$ is carried out using the fminbnd optimizer.
Our corresponding C++ implementation uses routines similar to those provided
in \cite{numrecipes2007}, section 16.1 (EM) and section 10.4 (minimization).

\section{Pre-Smoothed Particle Filters\label{sub:Pre-Smoothed-Particle-Filters}}
Equipped with an optimal one-step PS update, PSPF for accumulating the log-likelihood
$l=\log p(Y_{T})$ follows quite directly, and consists of the following steps:
\begin{enumerate}
\item Simulate $\{x_{0}^{(i),f}\}_{i=1}^{n}\sim p(x_{0})$, set $t=1$ and
$l=0$.
\item As for the SIR filter (section \ref{sub:The-SIR-filter}) set $x_{t}^{(i),p}=g(x_{t-1}^{(i),f},v_{t}^{(i)}),\; i=1,\dots,n$
to obtain an approximate sample from $p(x_{t}|Y_{t-1})$.
\item Compute optimal smoothing parameter $\bar b$ using the algorithm
given in section \ref{sec:bwimpl} with $\mathbf{x}=$$\{x_{t}^{(i),p}\}_{i=1}^{n}$, $y=y_{t}$
and $\mathcal M,\Sigma_\varepsilon$ given by the model specification. 
\item Update using the PS formulas (\ref{eq:py_estimate}-\ref{eq:pxcy_estimate})
with $\mathbf{x}=$$\{x_{t}^{(i),p}\}_{i=1}^{n}$, $y=y_{t}$ and $b=\bar b$. 
\item Sample $\{x_{t}^{(i),f}\}_{i=1}^{n}$ from the posterior representation
(\ref{eq:pxcy_estimate}) to obtain an equally weighted approximate
sample from $p(x_{t}|Y_{t})$.
\item Set $l\leftarrow l+\log(\frac{1}{n}\sum_{i=1}^{n}W_{i})$ so that
$l$ now approximates $\log p(Y_{t})$.
\item If $t<T$, set $t\rightarrow t+1$ and go to step 2, else stop.
\end{enumerate}
Various bias reduction techniques are available (see e.g. \citet{shepard_pitt97})
for computing the log-likelihood increment $\log p(y_{t}|Y_{t-1})\approx\log(\frac{1}{n}\sum_{i=1}^{n}W_{i})$,
but we do not employ those here in order to make comparisons easy. The PSPF
has, like the SIR filter computational complexity $O(Tn)$. We
have found that in practice the bottleneck in the PS update is actually
fitting the prior pilot $\hat{\pi}_{B}$ using the EM algorithm. Other,
more problem specific pilots are conceivable, but we do not discuss
this further here.

\subsection{\label{sec:Continuous-resampling}Continuous resampling }

The resampling step obtains equally weighted particles from a finite
mixture representation (\ref{eq:pxcy_estimate}) of the filter distribution.
Resampling is typically done by repetition of first drawing a component
in the mixture, and subsequently sampling from this component. This
process originates discontinuities in the simulated likelihood function,
even if common random numbers are applied for repeated evaluation,
and makes maximizing the simulated log-likelihood difficult. This
issue was first addressed by \citet{Pitt02smoothparticle}, who obtains
continuous draws from a univariate mixture of point masses. For multivariate
finite Gaussian mixture representations, \citet{Malik2011} provide
an algorithm that may be used to as the resampling step in the PSPF
for arbitrary $d_{x}$ (as the variance in each component of $\hat{p}(x_{t}|Y_{t})$ are equal). However for $d_{x}=1,2$ we have found algorithms
based on computing $\hat{p}(x_{t}|Y_{t})$ on a fine grid using fast
Fourier transform (FFT) methods and sampling from the corresponding
CDFs desirable (more detailed descriptions are
given in Appendix \ref{sec:cont-resample_app}). Like \citet{Malik2011}'s
algorithm, these FFT-based algorithms have linear complexity in $n$,
but we find them easier to program and tune.

\subsection{Comparison with other particle filters\label{sub:Comparison-with-other}}

To compare the proposed methodology with currently preferred particle
filters, we carry out some simulation experiments. 
\subsubsection{Experiment 1}
The first model we consider is
is given as 
\begin{eqnarray}
y_{t} & = & x_{t}+\varepsilon_{t},\;\varepsilon_{t}\sim N(0,\xi^{2}I_{d}),\; t=1,\dots,T,\label{eq:linmix_1}\\
x_{t} & = & 0.95x_{t-1}+\eta_{t},\nonumber \\
\eta_{t}& \sim & N(0,0.1\cdot \mathbf{1}_{d}+0.2I_{d}),\; t=1,\dots,T,\label{eq:linmix_2} 
\end{eqnarray}
where $\mathbf{1}_d$ denotes a $d\times d$ matrix with each element equal to 1.
The distribution of $x_0$ is a Gaussian mixture consisting of three equally weighted components
$N(0,I_{d})$, $N([1,\dots,1]',I_{d})$ and \\ $N([-1,1,-1,1,\dots]',I_{d})$. 
We consider each combination of dimensions $d=\{2,5,10\}$ and measurement error scales $\xi=\{0.01,0.1\}$,
The log-likelihood
$\log p(Y_{T})$ is available via the Kalman filter by conditioning
on each component in the $t=0$ mixture, and therefore admit comparison
between the particle filter-based log-likelihood approximations and
the truth for this globally non-Gaussian model. We consider a short
$(T=10)$ time series so that the non-Gaussian features introduced
by the initial Gaussian mixture do not die out. The contending filters
are PSPF, SIR, Ensemble Kalman Filter (EnKF), and post- and pre-smoothed 
regularized filters as described in \cite{musso01improving}. The latter
two are implemented using Gaussian kernels with bandwidth selection
based on the standard MISE-based plug-in formulas \citep[equation 2.3]{musso01improving},
and are therefore referred to as MISE-Post and MISE-Pre respectively.
The above mentioned filters rely
only on simulation from the state equation, and are therefore directly
comparable with respect to scope. As additional references, we also
compare with Auxiliary SIR (ASIR, based on the mean of $x_{t}|x_{t-1}$
as described in section 3.2 of \citet{pitt_shepard_1999}) and a fully
adapted Auxiliary SIR (FASIR, \citet{pitt_shepard_1999}). ASIR and
FASIR use knowledge of the mean of $x_{t}|x_{t-1}$, and knowledge
of the full distribution of $x_{t}|x_{t-1}$ respectively, and are
therefore not directly comparable to PSPF with respect to scope. We
report in Table \ref{tab:Monte-Carlo-estimates} the bias (loglike. bias), standard error 
(loglike. std.dev.)  and RMSE (loglike. RMSE) of the respective estimates
of $\log p(Y_{T})$ across 10,000 data sets simulated from (\ref{eq:linmix_1}-\ref{eq:linmix_2}).
In addition, as a crude measure for comparing
the posterior simulation performance, we also
report the square root of the expected squared Euclidian distance between
the mean of $\hat p(x_T|Y_T)$ and the simulated $x_T$ (filter RMSE).
We used $n=10,000$ for PSPF and $n=50,000$ for the other filters so
that the computing times for each filter are comparable when implemented
in MATLAB. The mean computing times relative to PSPF (relative CPU time) 
are also reported in Table \ref{tab:Monte-Carlo-estimates}.
For all filters but EnKF, non-continuous resampling was
performed in each time step.

From Table \ref{tab:Monte-Carlo-estimates}, we see that PSPF produces
smaller log-likelihood RMSEs then other filters that are based only on simulation
of the state, except in the $d=2,\xi=0.1$ case where MISE-Pre has the
smallest log-likelihood RMSE. However, assuming momentarily that
the log-likelihood RMSEs are $O(n^{-1/2})$, it should be noted that 
even in $d=2,\xi=0.1$ case, the log-likelihood RMSE would be the 
smallest for PSPF if the same $n$ was applied. 
The log-likelihood performances of the PSPF and the EnKF are fairly similar, but
it should be noted that the EnKF is not consistent, and therefore
the biases cannot be eliminated. For increasing dimensions and fixed
$n$, PSPF and EnKF becomes more similar, which is a consequence of
$\bar b$ being chosen closer to 0 in the high-$d$ cases (to counteract
the curse of dimensionality). 
SIR and MISE-Post perform poorly with respect to
log-likelihood estimation in all the high signal-to-noise
ratio ($\xi=0.01$) cases, and also in the moderate signal-to-noise
ratio ($\xi=0.1$) cases for $d=5,10$. MISE-Pre performs well
in the $d=2$ cases, but the performance relative to PSPF deteriorates
as $d$ grows. 

The ASIR exhibit highly variable
second stage weights, suggesting that the generic importance sampling
density implicitly introduced works poorly for this model. As it is
the optimal one step ahead particle filter, FASIR works extremely
well in all cases, with log-likelihood RMSEs that are two orders of magnitude smaller
than PSPF. Thus in the (not too often encountered) cases where full
adaptation is possible, one should opt for the FASIR over the
PSPF.

With respect to posterior simulation performance, PSPF produces filter RMSE results
almost identical to those of FASIR, indicating that the posterior samples of PSPF 
are close to those of the optimal one step ahead filter. 
MISE-Pre also produces filter RMSEs
close to those of FASIR, which underpins the claim made in 
Section \ref{sec:pycriterion} for this model, 
namely that the posterior estimator of pre-smoothed updates are
relatively insensitive to the choice of smoothing parameter. 
In the same vein, the log-likelihood results of PSPF relative to those of MISE-Pre
show that log-likelihood estimation is more sensitive to smoothing parameter selection
and therefore targeting MSE($\hat p(y)$) as is done here seems highly sensible.
\begin{table*}
\centering{}%
\begin{tabular}{lccccccc}
\hline 
 & PSPF & SIR & EnKF & MISE & MISE & ASIR & FASIR\tabularnewline
 &  &  &  & -Post & -Pre &  & \tabularnewline
\hline 
\multicolumn{8}{c}{$d=2,\;\xi=0.01$}\tabularnewline
\hline 
log-like. bias & -0.070 & -3.561 & -0.109 & -3.468 & -0.022 & -96.33 & 1.6e-5\tabularnewline
log-like. std. dev. & 0.303 & 27.49 & 0.418 & 24.82 & 0.328 & 27.55 & 0.006\tabularnewline
log-like. RMSE & 0.311 & 27.72 & 0.432 & 25.06 & 0.329 & 100.2 & 0.006\tabularnewline
filter RMSE & 0.014 & 0.017 & 0.014 & 0.017 & 0.014 & 0.017 & 0.014\tabularnewline
relative CPU time & 1.0  & 0.6 & 0.5 & 1.2 & 1.0  & 0.9  & 1.2\tabularnewline
\hline 
\multicolumn{8}{c}{$d=5,\;\xi=0.01$}\tabularnewline
\hline 
log-like. bias & -0.293 &  -1.6e3 & -0.356 & -1.6e3 & -0.572 & -1.5e3 & 9.0e-5\tabularnewline
log-like. std. dev. & 0.608 & 649.5 & 0.676 & 647.4 & 1.192 & 585.2 & 0.007\tabularnewline
log-like. RMSE & 0.675 & 1.8e3 & 0.764 & 1.8e3 & 1.322 & 1.6e3 & 0.007\tabularnewline
filter RMSE & 0.022 & 0.183 & 0.022 & 0.183 & 0.022 & 0.180 & 0.022\tabularnewline
relative CPU time & 1.0 & 1.1  & 1.2  & 2.4  & 1.7  & 1.6  & 2.1\tabularnewline
\hline 
\multicolumn{8}{c}{$d=10,\;\xi=0.01$}\tabularnewline
\hline 
log-like. bias & -0.612 & -2.2e4 & -0.634 & -2.3e4 & -4.086 & -2.1e4 & -3.8e-6\tabularnewline
log-like. std. dev. & 0.813 & 4.6e3 & 0.798 & 4.6e3 & 2.540 & 4.3e3 & 0.008\tabularnewline
log-like. RMSE & 1.018 & 2.3e4 & 1.019 & 2.3e4 & 4.811 & 2.2e4 & 0.008\tabularnewline
filter RMSE & 0.032 & 0.674 & 0.032 & 0.674 & 0.032 & 0.671 & 0.032\tabularnewline
relative CPU time & 1.0 & 1.4 & 1.7  & 2.9  & 2.1  & 2.1  & 2.6\tabularnewline
\hline 
\multicolumn{8}{c}{$d=2,\;\xi=0.1$}\tabularnewline
\hline 
log-like. bias & -0.066 & -0.024 & -0.104 & -0.019 & -0.013 & -16.53 & 3.5e-5\tabularnewline
log-like. std. dev. & 0.291 & 0.300 & 0.404 & 0.244 & 0.173 & 6.772 & 0.006\tabularnewline
log-like. RMSE & 0.299 & 0.301 & 0.417 & 0.245 & 0.174 & 17.86 & 0.006\tabularnewline
filter RMSE & 0.139 & 0.140 & 0.139 & 0.140 & 0.139 & 0.152 & 0.139\tabularnewline
relative CPU time & 1.0 & 0.8  & 0.5  & 1.5  & 1.0  & 1.0  & 1.1\tabularnewline
\hline 
\multicolumn{8}{c}{$d=5,\;\xi=0.1$}\tabularnewline
\hline 
log-like. bias & -0.278 & -3.423 & -0.340 & -3.381 & -0.304 & -50.47 & 3.8e-5\tabularnewline
log-like. std. dev. & 0.597 & 4.510 & 0.671 & 4.420 & 0.761 & 8.899 & 0.009\tabularnewline
log-like. RMSE & 0.658 & 5.662 & 0.752 & 5.564 & 0.819 & 51.25 & 0.009\tabularnewline
filter RMSE & 0.220 & 0.244 & 0.220 & 0.244 & 0.221 & 0.258 & 0.220\tabularnewline
relative CPU time & 1.0 & 1.1 & 1.2 & 2.2 & 1.7 & 1.7  & 2.0\tabularnewline
\hline 
\multicolumn{8}{c}{$d=10,\;\xi=0.1$}\tabularnewline
\hline 
log-like. bias & -0.584 & -131.2 & -0.611 & -131.6 & -2.694 & -129.2 & 7.6e-5\tabularnewline
log-like. std. dev. & 0.809 & 41.74 & 0.797 & 41.99 & 1.983 & 29.49 & 0.012\tabularnewline
log-like. RMSE & 0.998 & 137.7 & 1.005 & 138.1 & 3.345 & 132.5 & 0.012\tabularnewline
filter RMSE & 0.309 & 0.678 & 0.309 & 0.677 & 0.312 & 0.603 & 0.309\tabularnewline
relative CPU time & 1.0 & 1.4 & 1.7 & 2.7 & 2.1 & 2.1 & 2.6\tabularnewline
\hline 
\end{tabular}\caption{\label{tab:Monte-Carlo-estimates}Monte Carlo estimates of the log-likelihood
function for the model (\ref{eq:linmix_1}-\ref{eq:linmix_2}). All
quantities are calculated across 10,000 independent replica. The PSPF
is implemented with $n=10,000$ particles, whereas the other filters
are implemented with $n=50,000$ particles so that computing times
using MATLAB are on the same order.}
\end{table*}
\subsubsection{Experiment 2}
For the near-Gaussian model (\ref{eq:linmix_1}-\ref{eq:linmix_2}), 
the EnKF has a similar performance
to PSPF. To further explore the difference between PSPF and EnKF,
we consider a second simulation experiment that involves the non-linear
model 
\begin{eqnarray}
y_{t} & = & \frac{1}{20}x_{t}^{2}+\frac{1}{2}\eta_{t},\; t=1,\dots,T,\label{eq:kit_orig1}\\
x_{t} & = & \frac{1}{2}x_{t-1}+\sqrt{\frac{3}{4}}\varepsilon_{t},\; t=1,\dots,T,\label{eq:kit_orig2}\\
x_{0} & \sim & N(0,1),\label{eq:kit_orig3}
\end{eqnarray}
where $\eta_{t},\varepsilon_{t}\sim  N(0,1)$. 
The non-linear measurement equation is taken from a well-known
test case used by e.g. \cite{RSSB:RSSB736}. In particular, such models are
capable of generating bimodal filtering distributions as the sign
of $x_{t}$ cannot be determined from observations $y_{t}$. For the
PSPF and EnKF filters to be applicable, we need to augment the state
as indicated in section \ref{sub:The-updating-problem}, namely
\begin{eqnarray}
y_{t} & = & x_{t,1}+\frac{\sqrt{2}}{4}\eta_{t},\; t=1,\dots,T,\label{eq:kit_aug1}\\
x_{t,1} & = & \frac{1}{20}x_{t,2}^{2}+\frac{\sqrt{2}}{4}\varepsilon_{t,1},\; t=1,\dots,T,\label{eq:kit_aug2}\\
x_{t,2} & = & \frac{1}{2}x_{t-1,2}+\sqrt{\frac{3}{4}}\varepsilon_{t,2},\; t=1,\dots,T,\label{eq:kit_aug3}\\
x_{0} & \sim & N(0,1),\label{eq:kit_aug4}
\end{eqnarray}
where $\eta_{t},\varepsilon_{t,1},\varepsilon_{t,2} \sim N(0,1)$. The $N(0,1/4)$ observation noise in (\ref{eq:kit_orig1}) is
for simplicity split evenly between (\ref{eq:kit_aug1}) and (\ref{eq:kit_aug2}).
\begin{table*}
\centering{}
\begin{tabular}{lccccccc}
\hline 
Method & PSPF & EnKF & SIR &  & PSPF & EnKF & SIR\tabularnewline
\hline 
 & \multicolumn{3}{c}{$n_{PSPF}=10000$} &  & \multicolumn{3}{c}{$n_{PSPF}=50000$}\tabularnewline
\cline{2-4} \cline{6-8} 
log like. bias &  -0.188 & -0.437 & -0.094 &  &  -0.098 &  -0.437 &  -0.044\tabularnewline
log like. std. dev. & 0.815 & 1.426 & 0.605 &  & 0.546 &  1.426 & 0.387\tabularnewline
$q_{0.05}$ bias & 0.042 & 0.137 & 0.021 &  &  0.024 & 0.137 & 0.011\tabularnewline
$q_{0.05}$ std. dev. & 0.323 & 0.602 &  0.226 &  &  0.235 & 0.601 & 0.174\tabularnewline
$q_{0.2}$ bias & 0.051 &  0.120 & 0.025 &  &  0.033 &  0.120 &  0.014\tabularnewline
$q_{0.2}$ std. dev. & 0.396 &  0.464 &  0.350 &  & 0.321 &  0.464 & 0.264\tabularnewline
$q_{0.4}$ bias & 0.028 &  0.053 & 0.015 &  &  0.025 & 0.053 & 0.013\tabularnewline
$q_{0.4}$ std. dev. & 0.434 &  0.309 &  0.398 &  &  0.432 & 0.308 & 0.355\tabularnewline
relative CPU time &  1.0 &  0.4 &  0.8 &  &  1.0 & 0.8 &       1.9\tabularnewline
\hline 
\end{tabular}\protect\caption{\label{tab:sqr-mc}Monte Carlo estimates of 
log-likelihood and $p(x_{T,2}|Y_{T})$-quantiles for 
model (\ref{eq:kit_aug1}-\ref{eq:kit_aug4}) relative
to a reference SIR filter with 1,000,000 particles. PSPF was run with $n_{PSPF}$ particles
whereas EnKF and SIR were run with $5n_{PSPF}$ particles. The notation $q_{P}$
correspond to the $P$-quantile of $p(x_T|Y_T)=p(x_{T,2}|Y_{T})$. Due to the
symmetries of the model, the results for $q_{P}$, $P>1/2$ are essentially
equal to those for $q_{1-P}$ and are therefore not reported. All
computations where performed in MATLAB using 10,000 replications.
There is a factor 2.7 difference in the relative CPU times between $n_{PSPF}=10000$ and
$n_{PSPF}=50000$.}
\end{table*}

We are unaware of any computationally feasible exact method for calculating
$\log p(Y_{T})$ and $p(x_{T}|Y_{T})=p(x_{T,2}|Y_T)$ under either representation,
and therefore resort to a SIR filter with 1,000,000 particles applied
to representation (\ref{eq:kit_orig1}-\ref{eq:kit_orig3}) as the
reference. We choose $T=10$ and relatively low autocorrelation
and low signal to noise ratio to ensure that this reference method produces
reliable results. Specifically, repeated application of the reference
filter to a single simulated data set yields a standard error of the
log-likelihood estimate on the order of 0.001 and the standard error
of the estimated quantiles are on the order of 0.01 or better. The
setup of the simulation experiment is as follows. The reference method,
PSPF, EnKF and SIR (based on the representation in Equations 
\ref{eq:kit_aug1}-\ref{eq:kit_aug4})
where applied to 10,000 simulated data sets. We report bias and standard
deviation of the log-likelihood estimates relative to the reference
method. Further, for each of the contending filters we compare the
estimated $(0.05,0.2,0.4)$-quantiles of $p(x_T | Y_T)$ ($=p(x_{T,2} | Y_{T})$)
to the corresponding quantiles of the reference method, and report
bias and standard deviation. We consider two situations where the
PSPF has $n_{PSPF}={10000,50000}$ particles and the contending filters
have $5n_{PSPF}$ particles. This ensures that the filters
within each situation have similar computing times. The simulated data
sets $Y_{T}$ are the same in both situations. The results are reported
in Table \ref{tab:sqr-mc}. 

It is seen that the results for EnKF are close to identical when the
number of particles increases,
indicating that we incur substantial large-$n$ biases
by applying the EnKF to this model, in particular with respect to
log-likelihood evaluation and for the $(0.05,0.2)$-quantiles. This
is in contrast to PSPF, which as expected has diminishing biases and
standard deviations as $n_{PSPF}$ increases. Comparing PSPF to the SIR
with $n_{SIR}=5n_{PSPF}$ (columns 1 and 3), it is seen that PSPF and SIR provide comparable results for
comparable amounts of computing. Further, comparing PSPF in column 4
and SIR in column 3, where both filters have $n=50,000$, it is seen
that PSPF has a somewhat better log-likelihood performance 
whereas the filtering performance is roughly the same.
This indicates that $MSE(\hat p(y))$ is a sensible criterion for 
choosing the smoothing parameter both for the purpose of filtering
and likelihood evaluation, also for models with low signal to noise ratio.

\section{\label{sec:Illustrations}Illustrations}

Different aspects of PSPF are illustrated through two example models.
In section \ref{sub:One-factor-interest-rate} we consider a simple
non-linear interest rate model with high signal-to-noise ratio, under
which the PSPF is compared to other filters.
The second model (section \ref{sub:Example:-DSGE-model}) is included
to show that the PSPF can easily handle multiple states, and even
non-linear measurement equations via augmentation of the state vector. 

Throughout this section we refer to the quantity $l$ which is accumulated on step
7 of the PSPF algorithm (Section 4) as the simulated likelihood. Moreover
we refer to the maximizer of the simulated likelihood as the (off-line) simulated maximum
likelihood estimator along the lines of \cite{Malik2011}. Throughout both examples, the simulated maximum likelihood estimator is located using a BFGS numerical optimizer and finite difference 
gradients. Statistical standard errors are approximated using the (finite difference) observed information matrix at the optimizer. We prefer this approach over methods based on accumulating the score vector (and possibly on-line optimization) at each time step \citep{kantasetal09,1106.2525,Poyiadjis04022011} as
it is easier to program and adapt to new models.

\subsection{\label{sub:One-factor-interest-rate}One-factor interest rate model
with micro-structure noise}

The first example model we consider is the continuous time CEV diffusion,
\begin{equation}
dX_{\tau}=(\alpha-\beta X_{\tau})d\tau+\sigma\left(X_{\tau}\right)^{\gamma}dB_{\tau},\label{eq:CEV_ctime}
\end{equation}
of \citet{chan_et_al_1992}, where $B_{\tau}$ denotes a canonical
Brownian motion. We shall consider interest rate data available at
daily frequency, and a yearly time scale, corresponding to observations
at times $\tau=\Delta t,\; t=1,\dots,T$ where $\Delta=1/252$. We
apply an Euler-Maruyama discretization of (\ref{eq:CEV_ctime}), 
\begin{eqnarray}
x_{t} & = & x_{t-1}+\Delta(\alpha-\beta x_{t-1})+\sqrt{\Delta}\sigma x_{t-1}^{\gamma}\eta_{t},\label{eq:CEV_state}\\
\eta_{t} &\sim& iid\; N(0,1).\nonumber
\end{eqnarray}

It is well known that interest rate data are subject to micro structure
noise at daily frequency and a common workaround is to use data at
slower frequencies (see e.g. \citet{ait_sahalia_1999} who use monthly
data). To enable the usage of daily data, we model the micro structure
noise as being zero-mean Gaussian, i.e. 
\begin{equation}
y_{t}=x_{t}+\sigma_{y}\epsilon_{t},\;\epsilon_{t}\sim iid\; N(0,1).\label{eq:CEV_obser}
\end{equation}
Equations (\ref{eq:CEV_state}) and (\ref{eq:CEV_obser}) constitute
state-space systems on the form (\ref{eq:gen_SS_model_trans}-\ref{eq:gen_SS_model_obs})
with $\mathcal{M}=1$, $\Sigma_{\varepsilon}=\sigma_{y}^{2}$ and
parameter vector $\theta=(\alpha,\beta,\sigma,\gamma,\sigma_{y})$.
Thus we may estimate the parameters using PSPF-based simulated maximum
likelihood. 

The dataset considered is one-week nominal Norwegian Inter Bank Offered
Rate (NIBOR, in \%, $T=732$) between Jan. 2nd 2009 and Nov. 23rd
2011 obtained from the Norwegian central bank's website (http://www.norges-bank.no/).
Table \ref{tab:CEV-results}, lower panel, provides estimates and
statistical standard errors based on the observed Fisher information.
MC errors in both parameter estimates and standard deviation are evaluated
across 50 different seeds for the random number generator. We use
$n=2048$ as we aim for MC standard errors of the parameter estimates
on the order of 10\% of the statistical standard errors.
\begin{table*}
\centering{}
\small{\begin{tabular}{lcccccc} \hline 
 & $\log \alpha$ & $\log \beta$ & $\log \sigma $ & $\log \gamma $ &  $\log \sigma_y$ & log-like \\ \hline \noalign{\smallskip}
\multicolumn{7}{c}{$\sigma_y = 0$ } \\ 
& 2.084 & 1.314 & -0.970 & 0.104 &  & 1038.508\\ 
 & (0.122) & (0.123) & (0.013) & (0.015) &  & \\ 
\hline \noalign{\smallskip}
\multicolumn{7}{c}{$\sigma_y>0$ } \\ 
 & 1.570 & 0.815 & -1.612 & 0.486 & -3.739 & 1050.607\\ 
 & [0.022] & [0.023] & [0.017] & [0.009] & [0.013] & [1.229]\\ 
 & (0.389) & (0.397) & (0.176) & (0.114) & (0.100)\\ 
 & [0.007] & [0.008] & [0.001] & [0.001] & [0.003]\\ 
\hline 
\end{tabular} }\caption{\label{tab:CEV-results}Estimates and statistical standard errors
for the short term interest rate model (\ref{eq:CEV_state}-\ref{eq:CEV_obser})
applied to the NIBOR data. In the lower panel, estimates and statistical
standard errors are averaged across 50 estimation replica with different
seeds in the random number generator. Standard errors due to Monte
Carlo error for a single replica are presented in square parenthesizes
below the relevant figures.}
\end{table*}
\begin{figure}
\centering{}
\includegraphics[scale=0.4]{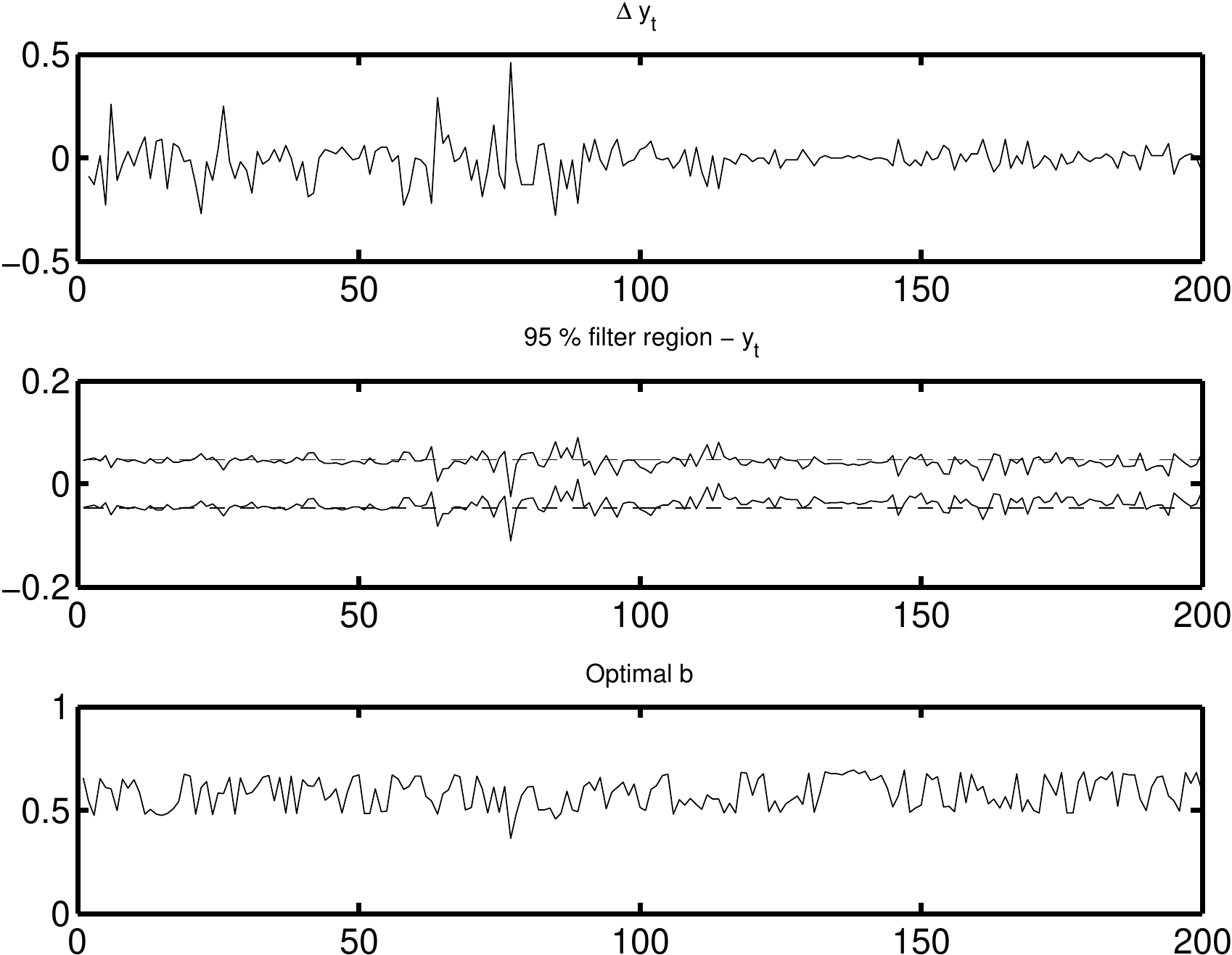}
\caption{\label{fig:Diagnostics-of-the}Diagnostics of the PSPF for 200 first
time steps under the interest rate model applied to the NIBOR data.
The upper panel displays returns $\Delta y_{t}\equiv y_{t}-y_{t-1}$.
In the middle panel, solid lines indicate the 95\% mass region of
the filter density with data subtracted. For comparison, dashed lines
indicate the estimated measurement error 95\% mass interval ($\pm1.96\exp(-3.739)$).
The lower panel plots the optimal smoothing parameters applied in
each update.}
\end{figure}
Typical computing times to maximize a simulated log-likelihood are approximately
$300$ seconds using a C++ implementation.
Four EM iterations based on all $n$ particles were employed, with
the EM computations distributed on 4 kernels of the 2010 laptop used
for calculations.

To contrast with not accounting for micro structure noise, we also
fitted the time-discretized CEV diffusion (\ref{eq:CEV_state}) directly
to the data using maximum likelihood, and report the results in the
upper panel of Table \ref{tab:CEV-results}. Judging from the log-likelihood
values, we find significantly better fit for the model accounting
for noise, and the estimates for the volatility structure, i.e. $\sigma$
and $\gamma$ are significantly different. As the PSPF does not require
the evaluation of transition probability densities $p(x_{t}|x_{t-1})$,
it is straight forward to apply more finely time-discretized versions
of (\ref{eq:CEV_ctime}) to the data. 
We found the single step discretization (\ref{eq:CEV_state}) to be
sufficiently accurate.

Figure \ref{fig:Diagnostics-of-the} provides some diagnostic plots
for $t=1,\dots,200$ and a randomly selected seed in the PSPF. There
are no signs of sample degeneracy, as the filter density is well spread
out during the whole time frame (and beyond). The fact that solid
and dashed lines almost overlap suggests that the model has a high
signal-to-noise ratio whereby most of the information in $p(x_{t}|Y_{t})$
originates from $p(x_{t}|y_{t})$. In the lower panel, it is seen
that the PS update is closer to the parametric update in cases with
large absolute returns $|\Delta y_{t}|$, whereas less smoothing is
imposed in easier cases corresponding to smaller returns.

As references for the PSPF, we also implemented FASIR, SIR and MISE-Pre
for this model. To avoid the complications associated with obtaining a continuous 
simulated log-likelihood for these filters \citep{Pitt02smoothparticle}, 
the reference filters where run 
with the parameters in Table \ref{tab:CEV-results}, lower panel, and the figures
reported below are across 100 replications.
The mean and MC standard error of the simulated log-likelihood for FASIR with $n=2048$
 reads 1051.5 and 0.52
respectively, showing that the PSPF is fully capable of competing
with specialized filters that exploit model-dependent structures.
The SIR filter with $n=65,536$  obtains an expected
log-likelihood 1016.9 and MC standard error of 7.3.
Thus allowing for the finite $n$ bias in the likelihood intrinsic
to the PSPF may be preferable over the highly variable but unbiased
particle filter. Finally, the MISE-Pre filter with $n=65,536$ yields expected log-likelihood
1020.4 with a MC standard error of 6.9. This highlights the need for
dynamic smoothing parameter selection for this model, even if pre-smoothing
with conjugate kernels is employed.

\subsection{\label{sub:Example:-DSGE-model}Dynamic stochastic general equilibrium
model}

A renewed interest in particle filters in the econometric
literature have at least partly been driven by the aim of estimating
non-linear solutions to dynamic stochastic general equilibrium (DSGE)
models \citep{fv_rr_2007,Amisano2010,andreasen_2011,dejong_unpub,flury_shephart_2011,Malik2011}.
We consider a simple neoclassical growth DSGE model \citep{King1988195,SchmittGrohe2004755},
with equilibrium condition given as
\begin{eqnarray*}
c_{t}^{-\gamma} & = & \beta E_{t}\left[c_{t+1}^{-\gamma}(\alpha A_{t+1}k_{t+1}^{\alpha-1}+1-\delta)\right],\\
c_{t}+k_{t+1} & = & A_{t}k_{t}^{\alpha}+(1-\delta)k_{t},\\
\log A_{t+1} & = & \rho\log A_{t}+\sigma_{A}\varepsilon_{t},
\end{eqnarray*}
where $c_{t}$ denotes optimal consumption, $k_{t}$ is capital and
$A_{t}$ is a positive productivity shock. A second order polynomial
approximation (replicating \citet{SchmittGrohe2004755}) to the solution
process in the log-deviation from non-stochastic steady states $\hat{c}_{t}=\log(c_{t}/\bar{c})$,
$\hat{k}_{t}=\log(k_{t}/\bar{k})$ and $\hat{A}_{t}=\log A_{t}$ is
applied. The resulting system may be written in state space form with
observation equation augmented with Gaussian noise
\begin{eqnarray}
\hat{c}_{t} & = & j(\hat{k}_{t},\hat{A}_{t})+\sigma_{c}\eta_{t,c},\label{eq:dsge_orig_obs}
\end{eqnarray}
and state evolution $\hat{k}_{t+1}=h(\hat{k}_{t},\hat{A}_{t})$, $\hat{A}_{t+1}=\rho\hat{A}_{t}+\sigma_{A}\eta_{t,A}$,
where $j,h$ are quadratic forms in their arguments and $\eta_{t,c},\eta_{t,A}\sim iid\; N(0,1)$. 

As the observation equation (\ref{eq:dsge_orig_obs}) is non-linear
in the state $(\hat{k}_{t},\hat{A}_{t})$, we use the augmentation-of-state
trick introduced in Section \ref{sub:The-updating-problem}. Specifically we include an additional
instrumental state $x_{t,3}$ to obtain a linear observation equation
and set in our notation $x_{t,1}=\hat{k}_{t}$, $x_{t,2}=\hat{A}_{t}$,
$y_{t}=\hat{c}_{t}$:
\begin{eqnarray}
y_{t} & = & x_{t,3}+r_{2}\sigma_{y}\eta_{t,y},\;\eta_{t,y}\sim iid\; N(0,1),\label{eq:dsge_filter_1}\\
x_{t,1} & = & h(x_{t-1,1},x_{t-1,2}),\label{eq:dsge_filter_11}\\
x_{t,2} & = & \rho x_{t-1,2}+\sigma_{A}\eta_{t,2},\;\eta_{t,2}\sim iid\; N(0,1),\label{eq:dsge_filter_111}\\
x_{t,3} & = & j(x_{t,1},x_{t,2})+r_{1}\sigma_{y}\eta_{t,3},\;\label{eq:dsge_filter_2}\\
\eta_{t,3}&\sim& iid\; N(0,1) \nonumber
\end{eqnarray}
where $r_{1},r_{2}>0$, conform to $r_{1}^{2}+r_{2}^{2}=1$. Thus
(\ref{eq:dsge_filter_1}-\ref{eq:dsge_filter_2}) conform with the
generic state space model (\ref{eq:gen_SS_model_trans}-\ref{eq:gen_SS_model_obs})
with $\mathcal{M}=[0\;0\;1]$ and $\Sigma_{\varepsilon}=r_{2}^{2}\sigma_{y}^{2}$.
In the computations, we fix $r_{1}$ to 0.05 to maintain some variation
in state $x_{t,3}$. We rely on a Maple script, called before each
run of the filter, to compute the second order approximation. As $x_{t,3}|Y_{t}$
is not used for prediction at time $t+1$, it suffices to use the
bivariate continuous resampling algorithm sketched in Appendix \ref{sec:cont-resample_app} for $x_{t+1,1},x_{t+1,2}|Y_{t}$. 

The structural parameters $\beta=0.95$, $\alpha=0.3$, $\gamma=2.0$
are kept fixed in simulation and estimation with values equal to those
considered in \citet{SchmittGrohe2004755} and the deprecation of
capital is kept at $\delta=0.5$. A simulated data set ($T=250$)
is generated with the remaining parameters $\theta=(\rho,\sigma,\sigma_{A})$
at $\text{logit}^{-1}(\rho)=2.0$, $\sigma_{A}=\sigma_{y}=0.1$ and
is subsequently subject to simulated maximum likelihood analysis using
the PSPF. 
\begin{table*}
\centering{}
\small{\begin{tabular}{ccccccccc} \hline \noalign{\smallskip}
  $\text{logit}^{-1}(\rho)$ & $\log(\sigma_y)$ & $\log(\sigma_A)$ & log-like & &
  $\text{logit}^{-1}(\rho)$ & $\log(\sigma_y)$ & $\log(\sigma_A)$ & log-like \\ \hline \noalign{\smallskip}
\multicolumn{4}{c}{$n=2048$} & & \multicolumn{4}{c}{$n=4096$} \\ 
\cline{1-4} \cline{6-9}
 1.89 & -2.73 & -2.13 & 128.11 & & 1.89 & -2.73 & -2.13 & 128.17\\  
 ${[}0.02{]}$ & ${[}0.01{]}$ & ${[}0.01{]}$ & ${[}0.46{]}$ & & ${[}0.01{]}$ & ${[}0.01{]}$ & ${[}$<0.01${]}$ & ${[}0.35{]}$\\ 
 (0.32) & (0.18) & (0.09) & & &  (0.33) & (0.19) & (0.09)\\ 
 ${[}$<0.01${]}$ & ${[}0.01{]}$ & ${[}$<0.01${]}$ & & &  ${[}$<0.01${]}$ & ${[}0.01{]}$ & ${[}$<0.01${]}$\\ 
\hline 
 \end{tabular} 
}\caption{\label{tab:DSGEtab}Estimates and statistical standard errors for
the DSGE model (\ref{eq:dsge_filter_1}-\ref{eq:dsge_filter_2}) based
on simulated data for different swarm sizes. Estimates and statistical
standard errors are averaged across 50 estimation replica with different
seeds in the random number generator. Standard errors due to MC error
for a single replica are presented in square parenthesizes below the
relevant figures.}
\end{table*}

Table \ref{tab:DSGEtab} provides parameter estimates and statistical
standard deviations, along with corresponding standard deviations
due to MC error across 50 independent replications of the experiments.
A typical computing time is 160 seconds to maximize a simulated log-likelihood
for $n=2048$ using our C++ implementation and 4 EM iterations distributed
on 4 kernels of the 2010 laptop used. We consider two different swarm
sizes, $n=2048$ and $n=4096$, to assess the robustness of the results,
and find only very small differences except for the obvious decreases
in MC uncertainties for the larger swarm. 

\citet{Malik2011} fit the same model using their (non-adapted) continuous
particle filtering method, but their routine required ``20000 particles
to obtain robust results'' for a somewhat shorter data set. They
do not report MC standard deviations or the values of structural parameters
they used, and therefore a direct comparison is difficult. However,
it is clear that the PSPF requires one order of magnitude fewer particles
to obtain robust results, which is very likely to justify the computational
overhead of the PSPF. 

As a further comparison we also implemented SIR and ASIR filters with and without
augmentation of the state, namely targeting either (\ref{eq:dsge_filter_1}-\ref{eq:dsge_filter_2}) or (\ref{eq:dsge_filter_11}-\ref{eq:dsge_filter_111}) along
with observation equation $y_t=j(x_{t,1},x_{t,2})+N(0,\sigma_y^2)$.
The parameters are kept fixed at values given in Table \ref{tab:DSGEtab}.
Firstly we find that the filters fares almost identically with
and without state augmentation, which suggest that state augmentation comes at
a small cost for this model.
Secondly, ASIR based on $E(x_{t+1}|x_t)$ fares much poorer than
the standard SIR filter, which indicates that the simple generic importance density
implied by this version of the ASIR has too thin tails and therefore more
model specific adaptation is in order. Finally, for the SIR
with (without) state augmentation we obtain a mean log-likelihood estimate reading
124.63 (124.61) and MC standard error in the log-likelihood estimate reading 0.38 (0.34) for $M=16384$ across 100 replications. Therefore the SIR MC standard errors are of the same order as the PSPF while using 4-8 times as many particles. 

We also 
observe that the log-likelihoods associated with the SIR are lower than the corresponding for PSPF, which indicate that the bias introduced by the PSPF is more
material here than for the previous example. There are several explanations for this, with the most prominent being that filtering distribution is less constrained
by the observation likelihood relative to the previous example, as $d_x>d_y$ in 
this case. This effect enables bias to build up over time. In addition, we observe that the kernel smoothing step of the PSPF introduces synthetic noise in the degenerate state transition (\ref{eq:dsge_filter_11}) and thereby implicitly
making the model more flexible, which may be contributing to the higher log-likelihood.

\section{\label{sec:Discussion}Discussion}

In this paper we explore the pre-smoothed update and the resulting
particle filter with a special emphasis on smoothing parameter selection.
Through simulation experiments and real data studies, the pre-smoothed particle filter
is shown to perform very well. In particular, we have shown that
the somewhat heuristic choice of one time period $MSE(\hat p(y))$ as the criterion
for choosing smoothing parameters also leads a competitive filter for many periods,
both in terms of log-likelihood evaluation and filtering.

The PSPF borrows ideas from a number of sources, including the filter
of \citet{alspach_sorenson_1972} and the subsequent literature, but
differ in the use of a resampling step. In \citet{alspach_sorenson_1972}
the mixture approximation of the posterior is propagated through the
system, allowing a non-uniform distribution of the weights to evolve. The
exact updating of finite Gaussian mixtures when the observation noise
is additively Gaussian is due to \citet{kotecha_djuric_2003}. More general
Pre-smoothed filters employing MISE-based smoothing parameter criteria that
are capable of handling more general observation equations via
a rejection sampling algorithm are discussed by 
\cite{leglandetal98,hurzeler1998,musso01improving,legland2004,1111.5866}. 
Shrunk kernel estimates in particle filters with constant smoothing parameter
were proposed by \citet{Liu2001}. The dynamic smoothing parameter
selection that we advocate is most closely related to that of 
\citet{Flury09learningand}, but their application was to a post-smoothed filter.
The effect smoothing parameter choice in pre-smoothed filters is also 
considered in \cite{legland2004} and \cite{1111.5866}, but they
target posterior simulation performance rather than likelihood performance.
The PSPF also borrows ideas from particle-based high-dimensional ($d_{x}$
large) data assimilation methods such as the Ensemble Kalman Filter
\citep{evensen_2003,rezaie_eidsvik_2012} in that Gaussian updating
formulas are used, but our focuses on potential applications and precision
are very different.

A major advantage of the proposed particle filtering approach is that
it is very easy to adapt to new models. The PSPF is not based on importance
sampling, and therefore the need for problem-specific importance densities,
and the potential for unbounded weight variance \citep{geweke_1989},
is mitigated. Provided an implementation of the PS update and smoothing
parameter selection (C++ and MATLAB source code is available from
the first author upon request), a user is only responsible for providing
routines for simulating the state equation and specifying the observation
equation. Implementation of the PS-update is also trivial when using
a high-level language such as MATLAB. Our implementation in MATLAB
used in section \ref{sub:Comparison-with-other} amounts to a few
dozen lines when using built in functions for minimizing $\bar C$
and fitting prior $\hat{\pi}_{B}$. 

One potential direction of further research is to assess the effect of the
choice of pilots $\hat \pi_B$, $\hat \pi_V$. In the present work, we have focused
on parametric pilots that lead to simple expressions for $f_0,f_1,f_2,f_3$, and that require
the least possible computational effort. However, any choices of finite Gaussian mixture
pilots, including fully non-parametric pilots \citep[see e.g.][]{wand_jones_94} with 
Gaussian kernels,
would lead to (more complicated) closed form expressions for $f_0,f_1,f_2,f_3$.
It would therefore be interesting to investigate whether adding more components
to the pilots would lead to substantially better results when 
the cost of more complicated computation associated with such an approach are taken into
account.

\bibliographystyle{chicago}
\bibliography{kleppe}

\appendix

\section{Calculations related to the practical MSE \label{sec:operational_calc}}
This section develops identity (\ref{eq:MSE_var_eq}) using iterative use of
the laws of total expectation and variance:
\begin{eqnarray*}
& & \underset{\tilde\Sigma,\tilde\mu,x^{(i)}\sim \text{ iid }\hat{\pi}_{V}}{Var}\left(\frac{1}{n}\sum_{i=1}^n \tilde W_i\right) \\
& =&\underset{\tilde\Sigma}{Var}\left(\underset{\tilde\mu,x^{(i)}\sim \text{ iid }\hat{\pi}_{V}}{E}
\left(\frac{1}{n}\sum_{i=1}^n \tilde W_i | \tilde\Sigma \right) \right) \\
& & + \underset{\tilde\Sigma}{E}\left(\underset{\tilde\mu,x^{(i)}\sim \text{ iid }\hat{\pi}_{V}}{Var}
\left(\frac{1}{n}\sum_{i=1}^n \tilde W_i | \tilde\Sigma \right) \right)
\end{eqnarray*}
\begin{eqnarray*}
& = & \underset{\tilde\Sigma}{Var}\left(
\underset{\tilde\mu}{E}\left(
\underset{x^{(i)}\sim \text{ iid }\hat{\pi}_{V}}{E}\left(
\frac{1}{n}\sum_{i=1}^n \tilde W_i |\tilde\Sigma, \tilde\mu \right)
|\tilde\Sigma \right) \right) \\
& &+\underset{\tilde\Sigma}{E}\left(
\underset{\tilde\mu}{Var}\left(
\underset{x^{(i)}\sim \text{ iid }\hat{\pi}_{V}}{E}\left(
\frac{1}{n}\sum_{i=1}^n \tilde W_i |\tilde\Sigma, \tilde\mu \right)
|\tilde\Sigma \right) \right) \\
& &+\underset{\tilde\Sigma}{E}\left(
\underset{\tilde\mu}{E}\left(
\underset{x^{(i)}\sim \text{ iid }\hat{\pi}_{V}}{Var}\left(
\frac{1}{n}\sum_{i=1}^n \tilde W_i |\tilde\Sigma, \tilde\mu \right)
|\tilde\Sigma \right) \right). \\
\end{eqnarray*}
Now the sums over the particles can be eliminated, so that the left hand side of (\ref{eq:MSE_var_eq})
can be written as
\begin{eqnarray*}
& &\underset{\tilde\Sigma}{Var}\left(
\underbrace{
\underset{\tilde\mu}{E}\left(
\underset{x^{(i)}\sim \hat{\pi}_{V}}{E}\left( W_i |\tilde\Sigma, \tilde\mu \right)
|\tilde\Sigma \right)}_{f_1(\tilde\Sigma)}
 \right) \\
& &+\underset{\tilde\Sigma}{E}\left(
\underset{\tilde\mu}{Var}\left(
\underset{x^{(i)}\sim \hat{\pi}_{V}}{E}\left(
\tilde W_i |\tilde\Sigma, \tilde\mu \right)
|\tilde\Sigma \right) \right) \\
& &+\frac{1}{n}\underset{\tilde\Sigma}{E}\left(
\underset{\tilde\mu}{E}\left(
\underset{x^{(i)}\sim \hat{\pi}_{V}}{Var}\left(
 \tilde W_i |\tilde\Sigma, \tilde\mu \right)
|\tilde\Sigma \right) \right). \\
\end{eqnarray*}
Finally, we use the $Var(X)=E(X^2)-(E(X))^2$ identity to obtain the desired expression: 
\begin{eqnarray*}
&&\underset{\tilde\Sigma}{Var}(f_1(\tilde\Sigma))\\
&&+\underset{\tilde\Sigma}{E}\left(
\underbrace{
\underset{\tilde\mu}{E}\left(
\left[\underset{x^{(i)}\sim \hat{\pi}_{V}}{E}\left(
 \tilde W_i |\tilde\Sigma, \tilde\mu \right)\right]^2
 |\tilde\Sigma \right)
 }_{f_3(\tilde\Sigma)}
  \right)\\
&&-\underset{\tilde\Sigma}{E}\left(\left[
\underbrace{
\underset{\tilde\mu}{E}\left(
\underset{x^{(i)}\sim \hat{\pi}_{V}}{E}\left(\tilde W_i |\tilde\Sigma, \tilde\mu \right)
|\tilde\Sigma \right)
}_{f_1(\tilde\Sigma)}
\right]^2\right)\\
&&+\frac{1}{n}\underset{\tilde\Sigma}{E}\left(
\underbrace{
\underset{\tilde\mu}{E}\left(
\underset{x^{(i)}\sim \hat{\pi}_{V}}{E}\left(
 \tilde W_i^2 |\tilde\Sigma, \tilde\mu \right)
|\tilde\Sigma \right)
}_{f_2(\tilde\Sigma)}
 \right) \\
&&-\frac{1}{n}\underset{\tilde\Sigma}{E}\left(
\underbrace{
\underset{\tilde\mu}{E}\left(\left[
\underset{x^{(i)}\sim \hat{\pi}_{V}}{E}\left(
 \tilde W_i |\tilde\Sigma, \tilde\mu \right)\right]^2
|\tilde\Sigma \right)
}_{f_3(\tilde\Sigma)} \right). \\
\end{eqnarray*}

\section{Continuous resampling details\label{sec:cont-resample_app}}

This section details algorithms for continuous (with respect to parameters)
sampling from a finite Gaussian mixture when the variance of each of the component
is equal. Both algorithms assume that the uniform random numbers applied
are the same for each run of the algorithm.

\subsection{Continuous resampling for $d_{x}=1$ }

In the case where the state is univariate, we have found that our
preferred re-sampling step consist of the following
\begin{itemize}
\item Grid: Compute mean and standard deviation of the FGM representation
of $p(x_{t}|Y_{t})$ and initiate a $n_{g}$-point regular grid containing
say the mean $\pm$ 8 standard deviations. Typical values of $n_{g}$
are 512 or 1024.
\item PDF: Noticing that the variance in each component in $\hat{p}(x_{t}|Y_{t})$
is equal, the PDF may be computed using fast Fourier transform methods
as explained thoroughly in \citet{Si86}, section 3.5 (with the modification
that each particle weight is now $w_{i}$ and not $1/n$).
\item CDF: Compute the cumulative distribution function (CDF) of the approximate
probability density function using a mid-point rule for each grid
point.
\item Fast inversion: Sample $\{x_{t}^{(i),f}\}$ based on stratified uniforms
using the CDF-inversion algorithm provided in Appendix A.3 of \citet{Malik2011}.
\end{itemize}
The total operation count of this algorithm is $O(n+n_{g}\log_{2}(n_{g}))$
and thus is linear complexity in the number of particles retained
also for continuous sampling.

\subsection{\label{sub:Continuous-resampling-2}Continuous resampling for $d_{x}=2$}

Also for $d_{x}=2$, it is possible to use fast Fourier transforms
to compute the posterior PDF at fine grid (say $n_{g,1}=n_{g,2}=256$)
by doing the smoothing in the Fourier domain. Our implementation rely
on the following steps
\begin{itemize}
\item Sample first dimension: Sample $\{x_{t,1}^{(i),f}\}$ from $p(x_{t,1}|Y_{t})$
using algorithm for $d_{x}=1$ (the marginal $p(x_{t,1}|Y_{t})$ is
easily recovered from the finite Gaussian mixture representation of $p(x_{t}|Y_{t})$). 
\item Joint PDF: Compute $\hat{p}(x_{t}|Y_{t})$ at a fine grid using 2-dimensional
fast Fourier transform methods and linear binning (see \citet{wand_1994}
for details).
\item Conditional CDFs: Compute the CDF of $x_{t,2}|x_{t,1}$ for each grid
point in the $x_{t,1}$-dimension from the joint PDF using a mid-point
rule.
\item Sample $x_{t,2}|x_{t,1}=x_{t,1}^{(i),f}$: For each $i=1,...,n$,
sample $x_{t,2}|x_{t,1}=x_{t,1}^{(i),f}$ using inversion sampling
based on a linear interpolation of the $x_{t,2}|x_{t,1}$-CDFs adjacent
to $x_{t,1}^{(i),f}$.
\end{itemize}
The computational complexity is \\ $O(n+n_{g,1}\log_{2}(n_{g,1})n_{g,2}\log_{2}(n_{g,2}))$,
i.e. linear in the number of particles.

\end{document}